\documentclass[sigconf]{acmart}
\AtBeginDocument{%
  }

\setcopyright{acmlicensed}
\copyrightyear{2026}
\acmYear{2026}
\setcopyright{cc}
\setcctype{by}
\acmConference[ICSE '26]{2026 IEEE/ACM 48th International Conference on Software Engineering}{April 12--18, 2026}{Rio de Janeiro, Brazil}
\acmBooktitle{2026 IEEE/ACM 48th International Conference on Software Engineering (ICSE '26), April 12--18, 2026, Rio de Janeiro, Brazil}
\acmPrice{}
\acmDOI{10.1145/3744916.3787784}
\acmISBN{979-8-4007-2025-3/2026/04}
\acmISBN{978-1-4503-XXXX-X/2018/06}

\usepackage{multirow}
\usepackage{amsmath}
\usepackage{amsfonts}
\usepackage{xcolor}
\usepackage{pifont}
\usepackage{subcaption}
\usepackage{framed}
\usepackage{listings}
\usepackage{booktabs}
\usepackage{caption}
\usepackage{graphicx}

\definecolor{backcolour}{rgb}{0.98,0.98,0.96}
\definecolor{codegray}{rgb}{0.5,0.5,0.5}
\definecolor{codeblue}{rgb}{0.0,0.2,0.6}
\definecolor{codepurple}{rgb}{0.58,0,0.82}

\lstdefinestyle{icsecode}{
  backgroundcolor=\color{backcolour},
  basicstyle=\ttfamily\scriptsize,
  commentstyle=\color{codegray},
  keywordstyle=\color{codeblue}\bfseries,
  stringstyle=\color{codepurple},
  breaklines=true,
  numbers=left,
  numberstyle=\tiny\color{gray},
  showstringspaces=false,
  captionpos=b
}
\definecolor{FindingBg}{gray}{0.95}   
\definecolor{FindingBar}{gray}{0.70}  
\newenvironment{findingbox}{%
  \par\smallskip
  \def\FrameCommand##1{%
    {\color{FindingBar}\vrule width 2pt}%
    \hspace{2pt}%
    \colorbox{FindingBg}{\hspace{4pt}##1\hspace{8pt}}%
  }%
  \MakeFramed{\advance\hsize-\width \FrameRestore}%
  \noindent\ignorespaces
}{%
  \endMakeFramed
  \par\smallskip
}

\definecolor{CodeBg}{gray}{0.95}
\definecolor{CodeBar}{gray}{0.70}

\lstdefinestyle{pycode}{
  language=Python,
  basicstyle=\ttfamily\small,
  columns=fullflexible,
  keepspaces=true,
  showstringspaces=false,
  breaklines=true,
  frame=none
}

\newenvironment{codebox}{%
  \par\smallskip
  \def\FrameCommand##1{%
    {\color{CodeBar}\vrule width 2pt}%
    \hspace{2pt}%
    \colorbox{CodeBg}{\hspace{12pt}##1\hspace{4pt}}%
  }%
  \MakeFramed{\advance\hsize-\width \FrameRestore}%
  \noindent\ignorespaces
}{%
  \endMakeFramed
  \par\smallskip
}
\newcommand{\cmark}{\textcolor{green!70!black}{\checkmark}} 
\newcommand{\xmark}{\textcolor{orange}{\ding{55}}} 

\newcommand{\todo}[1]{\textcolor{black}{ {#1}}}

\begin{document}

\title{Environment-Aware Code Generation: How far are We?}

\author{Tongtong Wu}
\authornote{These authors contributed equally to this work.}
\authornote{Corresponding author.}
\affiliation{
  \institution{Monash University}
  \city{Melbourne}
  \country{Australia}
}
\email{tongtong.wu@monash.edu}

\author{Rongyi Chen}
\authornotemark[1]
\affiliation{
  \institution{Southeast University}
  \city{Nanjing}
  \country{China}
}
\email{cry7632@gmail.com}

\author{Wenjie Du}
\affiliation{
  \institution{Southeast University}
  \city{Nanjing}
  \country{China}
}
\email{wenjedu0728@gmail.com}

\author{Suyu Ma}
\affiliation{
  \institution{CSIRO's Data61}
  \city{Melbourne}
  \country{Australia}
}
\email{suyu.ma@data61.csiro.au}

\author{Guilin Qi}
\affiliation{
  \institution{Southeast University}
  \city{Nanjing}
  \country{China}
}
\email{gqi@seu.edu.cn}

\author{Zhenchang Xing}
\affiliation{
  \institution{CSIRO's Data61}
  \city{Canberra}
  \country{Australia}
}
\email{zhenchang.xing@data61.csiro.au}

\author{Shahram Khadivi}
\affiliation{
  \institution{eBay Inc.}
  \city{Aachen}
  \country{Germany}
}
\email{skhadivi@ebay.com}

\author{Ramesh Periyathambi}
\affiliation{
  \institution{eBay Inc.}
  \city{San Francisco}
  \country{USA}
}
\email{rperiyathambi@ebay.com}

\author{Gholamreza Haffari}
\affiliation{
  \institution{Monash University}
  \city{Melbourne}
  \country{Australia}
}
\email{gholamreza.haffari@monash.edu}

\renewcommand{\shortauthors}{Tongtong Wu et al.}

\begin{abstract}Recent progress in large language models (LLMs) has led to impressive code generation capabilities. However, existing evaluations of LLMs primarily focus on generating isolated, small-scale code units (e.g., single functions or statements) under default or unspecified software environments. 
As a result, it remains unclear whether LLMs can reliably generate executable code tailored to specific user environments.
To fill this knowledge gap, we make the first systematic study of \textbf{Environment-Aware Code Generation (EACG)}, which requires generating code that is both functionally correct and directly executable under arbitrary software configurations. 
To support realistic evaluation, we introduce VersiBCB, a benchmark featuring \emph{multi-package}, \emph{executable-verified}, and \emph{deprecation-aware}, reflecting complex and evolving software environments that are often overlooked in prior datasets.
Building on this benchmark, we explore three orthogonal adaptation axes: \emph{data}, \emph{parameters}, and \emph{cache}, and further develop representative strategies for each.
Our results reveal that existing LLMs struggle with environment-specific code generation, but our adaptation strategies yield improvements in environment compatibility and executability.
These findings highlight critical challenges and opportunities for deploying LLMs in practical software engineering workflows.
\end{abstract}

\ccsdesc[500]{Software and its engineering~Software development techniques}
\ccsdesc[300]{Computing methodologies~Natural language processing}
\keywords{Large Language Models, Code Generation, Customization}


\maketitle

\section{Introduction}
\label{sect:intro}

Large Language Models (LLMs) have achieved significant progress in code generation, excelling at tasks such as auto-completion, test case synthesis, and API usage recommendation~\cite{jiang2024survey, chen2021evaluating, li2024testcase, chen2024evaluation, chen2024apigen}.
Despite their success, current LLM-based methods overlook a critical aspect of real-world software engineering: the variability of local development environments. 
In practice, developers work with a varied set of third-party libraries and versions\cite{libevolutioneval2025, wu2024versicode, wang2020libs} that often differ significantly from those encountered during model pretraining.
As a result, code generated by LLMs, while syntactically sound, frequently fails at runtime due to environment mismatches, such as incompatible APIs or deprecated functionalities. This limitation poses a fundamental barrier to the reliable deployment of LLMs in production workflows.

This problem becomes particularly severe when interacting with fast-evolving libraries. 
For example, the \texttt{transformers} library has introduced substantial changes across versions, including renamed function arguments and reorganized module paths, which can cause previously working code to break or behave unexpectedly \cite{ transformers_migration_v4}.
Real-world development becomes even more complex when multiple libraries like \texttt{torch}, \texttt{transformers}, and \texttt{datasets} must interoperate with strict version constraints and dependency relationships.
These challenges become even more pronounced in collaborative or legacy systems, where environments are often fragmented and difficult to standardize \cite{assuncao2024modernization}.
In these settings, teams must manage outdated dependencies, sparse documentation, and entrenched system complexity, making upgrades risky and costly.
Such scenarios highlight the importance of \emph{environment-aware} code generation, where LLMs are able to produce code that is tailored to the user’s specific software configuration.


Inspired by the modularity of modern development tools, we pose the central research question: \emph{Can LLMs be configured to respect local software environments, much like how developers configure IDEs or virtual environments?}
While tools like \texttt{requirements.txt} and \texttt{environment.yml} help manage package dependencies, LLMs are typically deployed as static, one-size-fits-all systems. Customizing their behavior to reflect a user's package landscape, without retraining for every possible configuration, remains a major technical challenge.
To formalize this setting, we introduce the task of \textbf{Environment-Aware Code Generation (EACG)}: given a functional requirement and an explicit environment specification (i.e., the set of installed packages and their versions), generate code that is not only functionally correct but also executable in the specified environment. Addressing EACG requires models to reason over both high-level intent and low-level library semantics, while remaining compositional across diverse and potentially unseen package combinations.

To systematically evaluate the EACG task, we construct \textbf{VersiBCB}, a dedicated benchmark grounded in real-world software development scenarios. 
VersiBCB is built by collecting code examples along with environment specifications from a diverse set of open-source Python projects. 
For each task, we curate the functional requirements, such as problem descriptions or usage examples, and specify the exact versions of all relevant packages. 
This approach captures a wide spectrum of library combinations and version constraints that are commonly encountered in practice. 
As a result, VersiBCB reflects both widely used environments and rare edge cases involving rapidly evolving or seldom combined libraries. 
Each instance in the benchmark is validated for executability by running the generated code in a controlled environment that matches the specified configuration. 

To tackle the challenges posed by environment-aware code generation, we investigate three complementary strategies for customizing LLM behavior at inference time, each grounded in a different adaptation principle: 
\textbf{Data-based} adaptation via \textit{retrieval-augmented generation (RAG)}: injecting environment-relevant documentation or code examples into the context window.
\textbf{Parameter-based} adaptation via \textit{modular fine-tuning with \todo{mixture-of-experts (MoE)}}: activating specialized submodules aligned with target library versions.
\textbf{Cache-based} adaptation via \textit{memory-augmented generation}: using a plugin-style key-value memory to incorporate environment-specific guidance at inference.
\noindent Each method reflects a different customization axis, i.e., data, parameters, and memory, providing a comprehensive view of how LLMs can be made environment-aware.
%

Our evaluation reveals that, while adaptation strategies such as retrieval-augmented generation, memory augmentation, and mixture-of-experts can improve LLM performance in environment-aware code generation,  limitations still persist. 
Memory-based approaches excel in code migration tasks by leveraging cached environment-specific patterns but are prone to errors when cached knowledge does not precisely align with new requirements. 
Retrieval-augmented generation can incorporate relevant context but often overfits to retrieved examples and fails to reconcile subtle semantic differences.
Mixture-of-experts methods offer robust compatibility with explicit version constraints but can suffer from expert misrouting, particularly in novel or distribution-shifted environments.
Across all strategies, performance degrades markedly in complex domains like machine learning and in scenarios involving unseen library and version combinations. 
%
In summary, this paper makes the following contributions:
\begin{itemize}
    \item We introduce and formalize the task of \textbf{Environment-Aware Code Generation (EACG)}, and present \textbf{VersiBCB}, the first large-scale benchmark for this task, built from real-world Python projects with diverse environment specifications.

    \item We are the first to investigate the ability of LLMs in environ-\\ment-aware code generation, systematically evaluating three adaptation strategies across two representative tasks.
    \item We report key findings and implications on LLM capability, and discuss future directions for robust environment-aware code generation.
\end{itemize}

\section{Motivation and Problem Formulation}
\label{sect:problem}

Although LLMs demonstrate strong capabilities on standard code generation benchmarks~\cite{yu2024codereval,du2024evaluating,liu2023your,lai2023ds,li2022competition,chen2021evaluating,austin2021program,athiwaratkun2022multi,hendrycks2021measuring}, their outputs frequently encounter failures when deployed in actual development environments~\cite{jimenez2023swe,yang2024swe}. 
A major challenge arises when generated code must operate with diverse combinations of third-party libraries and specific package versions, as commonly found in modern software projects. 
Incompatibilities or version mismatches can cause otherwise correct code to fail at runtime, requiring developers to spend additional effort resolving subtle dependency issues. 


Consider the \texttt{transformers} library as a representative example. Code generated for version \texttt{4.36.2} may rely on methods such as \texttt{model.push\_to\_hub()}, which are either unavailable or semantically different in earlier versions like \texttt{4.8.0}. Seemingly trivial changes, including renamed parameters, modified class constructors, or refactored import paths, can lead to non-trivial runtime failures. These challenges are exacerbated in multi-library contexts, where transitive dependencies and version entanglement further complicate compatibility. For instance, using \texttt{torch==1.9.0}, \texttt{transformers==4.8.0}, and \texttt{datasets==1.12.0} concurrently imposes non-trivial constraints on compositional correctness across libraries, constraints that current LLMs are ill-equipped to handle.
Such cases highlight a fundamental limitation: unlike human developers who actively manage dependency versions to ensure compatibility, current LLMs operate under static assumptions and lack awareness of the target execution environment. In real-world development workflows, dependency configurations are often non-default, outdated, or highly customized, reflecting substantial heterogeneity across systems. This underscores the need for \textit{environment-aware LLMs}, models capable of conditioning code generation on user-specific dependency contexts to ensure functional correctness and practical usability.

\subsection{Problem Formulation}
\label{sec:problem_formulation}

\begin{figure}[t]
    \centering
    \includegraphics[width=.95\linewidth]{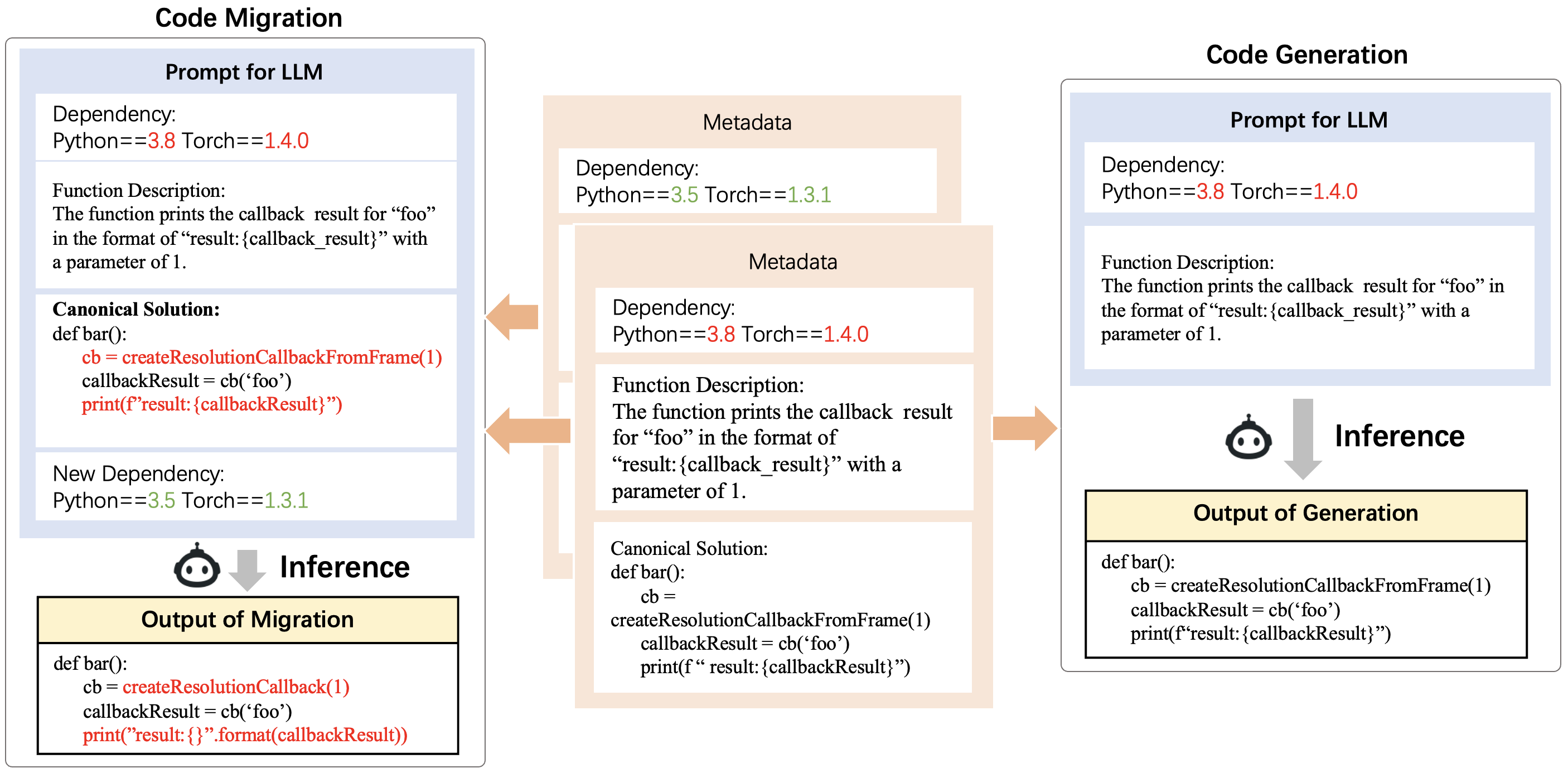}
    \caption{Task definition covering both {environment-aware} code generation and code migration.}
    \Description{Task definition covering both {environment-aware} code generation and code migration.}
    \label{fig:task_definition}
\end{figure}


{
We formalize the task of \textbf{Environment-Aware Code Generation (EACG)} as a conditional code generation problem under environment constraints. Each meta instance is defined as $m = [L, V, d, t] \in \mathcal{M}$, 
where $L = [l_1, l_2, \ldots, l_n]$ is the list of dependent packages, $V = [v_1, v_2, \ldots, v_n]$ represents the corresponding package versions, $d$ is a functional requirement expressed in natural language, $t: \mathcal{C} \to \{\texttt{True}, \texttt{False}\}$ is a test function that verifies whether a generated code snippet $C$ satisfies $d$ in the environment $(L, V)$.
}
%
{
As shown in Fig.~\ref{fig:task_definition}, we further extend EACG to a code migration setting, referred to as Environment-Aware Code Migration (EACM), to simulate real-world software maintenance scenarios.
Given a pair of source environment $(L, V_{\text{src}})$ and target environment $(L, V_{\text{tgt}})$ with the same requirement $d$, the goal is to migrate a valid code snippet $C_{\text{src}}$ into $C_{\text{tgt}}$.
}
{The above formulations capture} two essential capabilities: generating executable code that conforms to version constraints, and adapting code to a new environment while preserving intended functionality.
We evaluate model performance using three protocols:

\begin{findingbox}\small
\noindent\textbf{Executability:} whether the generated code executes successfully under the given environment;

\noindent\textbf{Compatibility:} whether all API calls are valid under the specified package versions;

\noindent\textbf{Composability:} whether the model generalizes to unseen combinations of packages and versions.
\end{findingbox}

\section{VersiBCB}
\label{sect:dataset}

To support research on environment-aware code generation, we construct \textbf{VersiBCB}, a large-scale benchmark specifically designed to reflect real-world software environments. As shown in Table~\ref{tab:dataset-summary}, VersiBCB fills critical gaps left by existing datasets. Unlike prior benchmarks that are either synthetic, LLM-generated, or restricted to isolated libraries, VersiBCB provides \emph{expert-curated}, \emph{execution-verified}, and \emph{multi-library} migration examples grounded in real Python projects. It is the first to systematically support \textbf{cross-package evolution} and \textbf{deep multi-step version transitions}, enabling evaluation under highly realistic and complex development settings.

\begin{table}[ht]
\centering
\caption{
Comparison of version-aware code datasets along key dimensions. 
\textbf{Problems}: number of code migration instances. 
\textbf{Source}: origin of the data (e.g., GitHub, LLM-generated, expert-curated).
\textbf{Lib.}: library-specific evolution.
\textbf{Exec.}: execution-based validation. 
\textbf{Real.}: real-world (non-synthetic) code.
\textbf{Cross.}: presence of cross-package evolution. 
\textbf{Deep.}: support for in-depth multi-step evolution.
}
\smallskip
\resizebox{\linewidth}{!}{
\begin{tabular}{lrrrrrrr}
\toprule
\textbf{Dataset} & \textbf{Problems} & \textbf{Source} & \textbf{Lib.} & \textbf{Exec.} & \textbf{Real.} & \textbf{Cross.} & \textbf{Deep.} \\
\midrule
Versicode         & 98,692  & GitHub, SO      & \cmark & \xmark & \cmark & \xmark & \xmark \\
CodeUpdateArena   & 670     & LLM-Gen         & \cmark & \cmark & \xmark & \xmark & \xmark \\
Wang et al.       & 28,125  & API Logs        & \cmark & \xmark & \cmark & \xmark & \xmark \\
GitChameleon      & 116     & Manual + LLM    & \cmark & \cmark & \cmark & \xmark & \xmark \\
\midrule
versiBCB          & 725     & Expert          & \cmark & \cmark & \cmark & \cmark & \cmark \\
\bottomrule
\end{tabular}
}
\label{tab:dataset-summary}
\end{table}

\subsection{Dataset Construction}

\begin{figure}[t]
    \centering
    \includegraphics[width=\linewidth]{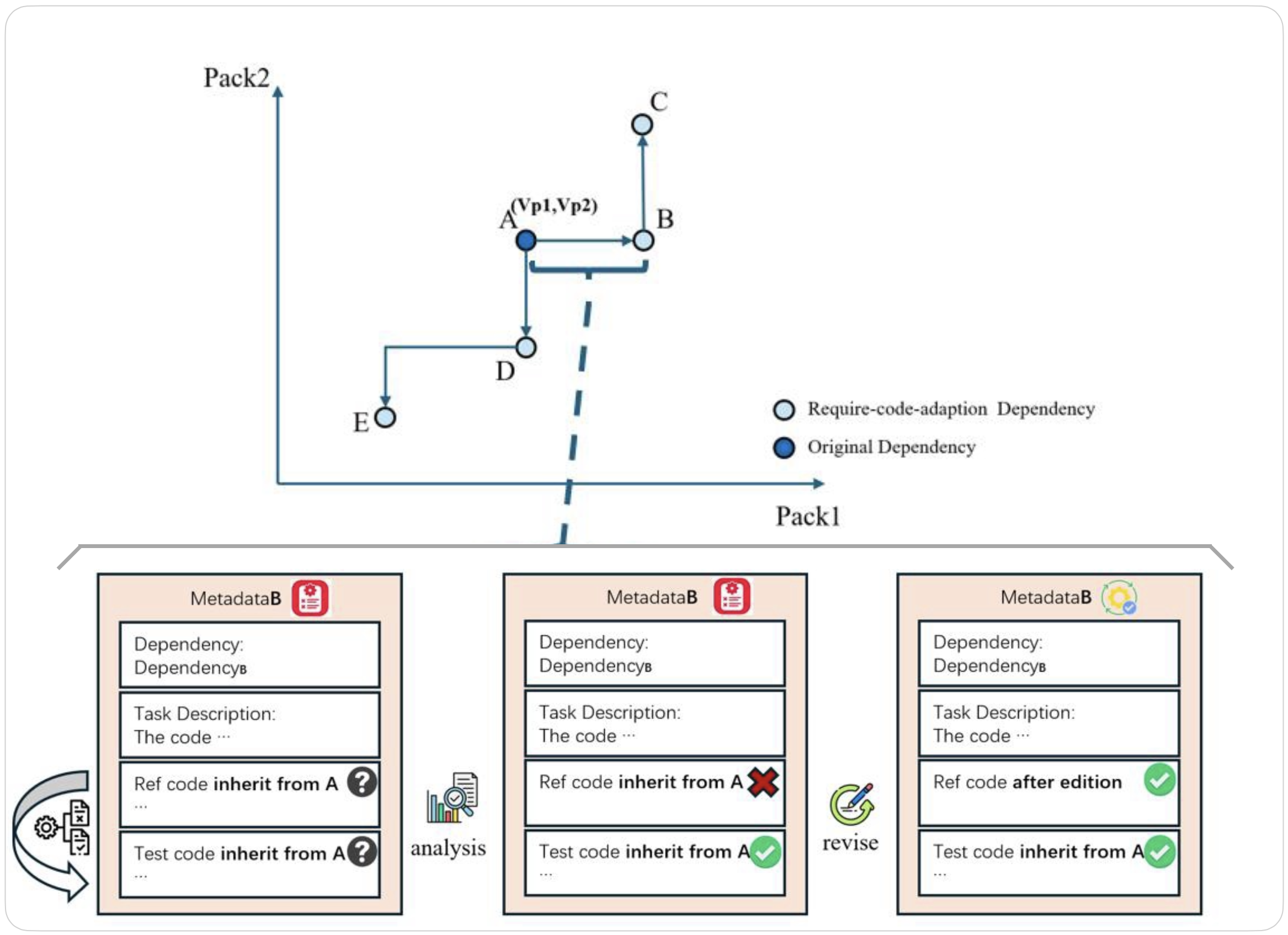}
    \caption{Overview of dataset construction via bidirectional environment traversal.}
    \Description{Overview of dataset construction via bidirectional environment traversal.}
    \label{fig:data_construction}
\end{figure}

VersiBCB is derived from BigCodeBench\footnote{{\url{https://bigcode-bench.github.io/}}}~\cite{abs-2406-15877} by augmenting its metadata instances with environment-aware annotations. For each metadata entry, we first determine a set of feasible dependency combinations. Starting from a known executable configuration (point A in Figure~\ref{fig:data_construction}), we simulate both forward and backward version transitions across multiple libraries. 
{To ensure the reasonableness of package version combinations in VersiBCB, we automatically create an isolated Python environment for each sample using Anaconda. If a specified version combination is not feasible, such as when dependencies conflict, the environment fails to build and the sample is excluded. }
Such bidirectional traversal identifies evolution points where the reference code fails due to API deprecation, signature changes, or behavioral divergence.
To localize these failures and produce corrected variants, we employ a reflection-guided LLM-based agent that leverages source code, documentation, and test feedback. The agent automates code repair by detecting obsolete API calls, rewriting affected lines, and updating associated test cases. This process enables precise annotation of version-sensitive migration tasks under the EACG and EACM protocols.

\subsection{Dataset Statistics}
Table~\ref{tab:dataset_stats} presents the key statistics of VersiBCB. The dataset comprises 725 executable instances spanning two core tasks: code generation and code migration, each further split by whether deprecated APIs are permitted in the target outputs. 
The \textit{Deprecation Policy} column differentiates between two evaluation settings: \textit{Lenient}, which permits the use of deprecated APIs if still functional, and \textit{Strict}, which requires exclusive use of actively supported APIs. Instances under the \textit{Strict} policy are selected based on source-level deprecation alerts, enabling precise analysis of model compliance with up-to-date API standards.
This design enables controlled evaluation under both permissive and strict deprecation policies, reflecting common real-world maintenance scenarios.
VersiBCB covers 36 widely used Python libraries and 202 distinct versions, resulting in 601 unique environment configurations. These configurations are derived from real-world projects and curated to capture meaningful semantic shifts such as API signature changes and behavioral divergence. Unlike prior benchmarks that operate in static or synthetic settings, VersiBCB emphasizes cross-library and cross-version compatibility, simulating the complexity of modern software environments.
The dataset is balanced across task types. Generation tasks include 417 instances, while migration tasks include 416. On average, migration inputs are longer (950–971 tokens) than generation inputs (830–849 tokens), reflecting the additional reasoning required to adapt code across environments.

\begin{table}[t]
\centering
\caption{Dataset statistics under different deprecation policies for code generation and migration.}
\label{tab:dataset_stats}
\renewcommand{\arraystretch}{1.2}
\resizebox{0.9\linewidth}{!}{
\begin{tabular}{l|cc|cc}
\toprule
\textbf{Property} & \multicolumn{2}{c|}{\textbf{Code Generation}} & \multicolumn{2}{c}{\textbf{Code Migration}} \\
\cmidrule(lr){2-3} \cmidrule(lr){4-5}
& \textbf{Lenient} & \textbf{Strict} & \textbf{Lenient} & \textbf{Strict} \\
\midrule
Avg. Input Tokens & 830 & 849 & 950 & 971 \\
Num. of Instances & 335 & 82  & 334 & 82 \\
\bottomrule
\end{tabular}
}
\end{table}

\subsection{Domain Distribution}

\begin{figure}
    \centering
    \includegraphics[width=\linewidth]{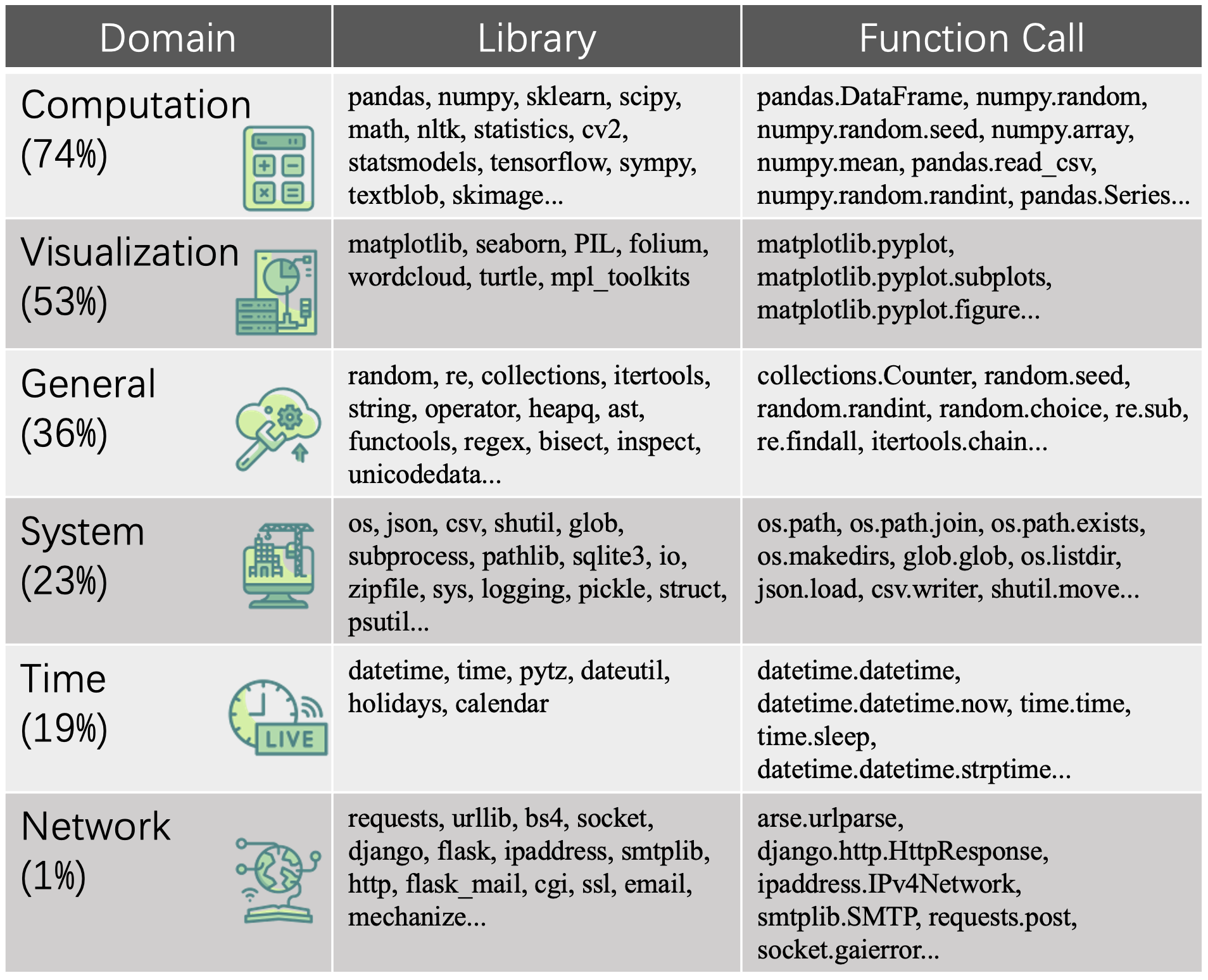}
    \caption{Distribution of VersiBCB tasks across domains, based on the presence of domain-specific libraries.}
    \Description{Distribution of VersiBCB tasks across domains, based on the presence of domain-specific libraries.}
    \label{fig:data_distribution}
\end{figure}

VersiBCB is designed to capture realistic and diverse Python development environments. These environments span both common and edge case scenarios across major domains such as machine learning, data processing, visualization, and scientific computing~\footnote{{To assess ecosystem relevance, we collected GitHub star counts for representative libraries, as of 26~Oct~2025, including NumPy (30.5k), Pandas (46.6k), scikit learn (63.4k), SciPy (14k), Seaborn (13.5k), Requests (53.3k), urllib (3.9k), and Flask (70.4k). The high popularity of these libraries reflects their widespread adoption and indicates that VersiBCB covers high impact libraries frequently used in practice.}}. 
To further examine the generalizability of VersiBCB, we analyze its distribution across domains such as computation, data processing, machine learning, and visualization (Figure~\ref{fig:data_distribution}). Each domain’s proportion reflects the frequency of tasks involving domain specific libraries. For example, the “Computation” domain accounts for 74\% of tasks, suggesting that most instances involve at least one numerical or scientific computing library. This distribution demonstrates that VersiBCB not only emphasizes ecosystem realism but also maintains balanced coverage across widely used domains.

\subsection{Dataset Analysis}


We evaluate several state-of-the-art LLMs on VersiBCB to assess the difficulty of the dataset and the gap between current model capabilities and the demands of environment aware code generation. 
{Model selection was guided by popularity, diversity, and accessibility. We included widely used open source code LLMs (e.g., \texttt{CodeLlama}, \texttt{StarCoder}, \texttt{DeepSeek Coder}, \texttt{CodeGemma}) to ensure reproducibility and representativeness of community baselines. In addition, we evaluated leading commercial models (e.g., \texttt{GPT 4.1}, \texttt{Gemini}) as reference points for top-tier performance. The selected models span different parameter scales and training paradigms, ensuring that our conclusions capture general trends across the model landscape rather than reflecting the behavior of a single model family.
}
Table~\ref{tab:model_performance} reports Pass@k results on both code generation (EACG) and migration (EACM) tasks. 

{
The \textit{Pass@k} metric evaluates functional correctness in code generation, measuring the fraction of tasks for which at least one of the top-$k$ generated programs passes all unit tests:
\begin{align}
\text{Pass@}k = \frac{1}{N} \sum_{i=1}^{N} \mathbb{I}\!\left[\exists\, j \le k : 
\text{AllTestsPass}(C_{i,j}) \right],
\end{align}
where $\mathbb{I}[\cdot]$ is the indicator function. A higher value indicates a greater likelihood of producing at least one correct solution among the top-$k$ candidates.
}
Results show that code generation under version constraints is particularly challenging for small to medium-sized models, e.g., \texttt{CodeGemma-7B} ~\cite{team2024codegemma} and \texttt{CodeLlama-13B}~\cite{roziere2023code} achieve Pass@1 rates below 1\%, while larger models such as \texttt{Deep-\\Seek-v3}~\cite{liu2024deepseek} and \texttt{GPT-4.1-mini}~\cite{openaiOpenAIPlatform} achieve moderate success. Notably, all models perform better on the code migration task, indicating that adapting from an existing code context is easier than generating from scratch under environment constraints.

\begin{table}[ht]
\centering
\caption{
Performance of various models on code generation and migration tasks. 
Model names are abbreviated: DS-7B (\texttt{DeepSeek-Coder-7B}), CG-7B (\texttt{CodeGemma-7B-it}), CL-13B (\texttt{CodeLlama-13B}), 
SC2-15B (\texttt{StarCoder2-15B}~\cite{lozhkov2024starcoder}), LL3-70B (\texttt{LLaMA3-70B}~\cite{dubey2024llama}), GPT4.1 (\texttt{GPT-4.1-mini})~\cite{openaiIntroducingGPT41}, DS-v3 (\texttt{DeepSeek-v3}).
Metrics are reported as Pass@k (\%).
}
\label{tab:model_performance}
\resizebox{\linewidth}{!}{
\begin{tabular}{@{}lcccccc@{}}
\toprule
\multirow{2}{*}{Model} & \multicolumn{3}{c}{Code Generation} & \multicolumn{3}{c}{Code Migration} \\ \cmidrule(l){2-7} 
                       & Pass@1 & Pass@3 & Pass@5            & Pass@1 & Pass@3 & Pass@5            \\ \midrule
DS-7B                 & 0.00   & 0.00   & 0.00              & 2.99   & 6.59   & 8.38              \\
CG-7B                 & 0.60   & 0.90   & 2.69              & 13.47  & 34.13  & 49.70             \\
CL-13B                & 0.30   & 0.90   & 1.79              & 18.26  & 36.23  & 49.10             \\
SC2-15B               & 0.00   & 0.00   & 0.30              & 5.99   & 15.87  & 21.26             \\
LL3-70B               & 18.51  & 24.78  & 27.76             & 57.19  & 60.78  & 61.98             \\
GPT4.1                    & 27.76 & 32.24 & 33.43 & 53.29 & 59.28 & 61.68 \\
DS-v3                 & 23.88  & 28.06  & 30.75             & 66.17  & 70.06  & 70.66             \\ \bottomrule
\end{tabular}}
\end{table}



{
To validate the quality and uniqueness of VersiBCB, we selected benchmarks that are representative, complementary, and comparable. Table~\ref{tab:benchmark_comparison} contrasts Pass@1 performance on VersiBCB with standard benchmarks such as \texttt{HumanEval+}~\cite{evalplus} and \texttt{LiveCodeBench}~\cite{jain2024livecodebench}. \texttt{HumanEval+} serves as a widely adopted baseline for function level code generation, reflecting generic code synthesis ability. \texttt{LiveCode\\Bench} introduces execution verified tasks at a larger scale, providing a more challenging yet still environment agnostic setting. In contrast, VersiBCB explicitly targets environment aware code generation and migration across multiple libraries, versions, and API deprecations, with execution based validation. 
}
While models achieve over 80\% on \texttt{HumanEval+} and over 40\% on \texttt{LiveCodeBench}, their performance drops sharply on VersiBCB, especially for the EACG task. This suggests that environment constraints significantly increase task complexity and expose limitations in existing LLMs.

\begin{table}[t]
\centering
\caption{
Pass@1 performance comparison across standard benchmarks and our VersiBCB dataset. 
Model names are abbreviated as follows: 
{G4o} (\texttt{gpt-4o-mini})~\cite{openaiGPT4oMini}, 
{G4.1} (\texttt{gpt-4.1-nano})~\cite{gpt-4.1-nano}, 
{Gemini} (\texttt{Gemini-1.5-flash-002})~\cite{team2023gemini}, 
{G3.5} (\texttt{gpt-3.5-turbo}), and
{DS-v3} (\texttt{DeepSeek-v3}).
}
\label{tab:benchmark_comparison}
\resizebox{\linewidth}{!}{
\begin{tabular}{@{}lccccc@{}}
\toprule
\multirow{2}{*}{\textbf{Benchmark}} & \multicolumn{5}{c}{\textbf{Model (Pass@1)}} \\ \cmidrule(l){2-6} 
                                    & \texttt{G4o} & \texttt{G4.1} & \texttt{Gemini} & \texttt{G3.5} & \texttt{DS-v3} \\ \midrule
HumanEval+                         & 83.5         & 82.9          & 79.3            & 69.5          & 88.4           \\
LiveCodeBench (CodeGen)            & 35.5         & 42.0          & 36.1            & 22.8          & 49.6           \\ \midrule
VersiBCB-CM (Code Migration)       & 35.6         & 51.8          & 55.7            & 54.5          & 66.2           \\
VersiBCB-CG (Code Generation)      & 20.6         & 19.4          & 21.5            & 13.1          & 23.9           \\ \bottomrule
\end{tabular}}
\end{table}

To further understand this discrepancy, we compute Pearson correlations between benchmark scores and VersiBCB results in Table~\ref{tab:correlation_analysis}. We find a strong correlation between existing benchmarks and VersiBCB-CG (e.g. $r = 0.925$ with \texttt{HumanEval+}), but a much weaker correlation with VersiBCB-CM (e.g. $r = 0.095$). 
{
While LLMs perform slightly better on migration tasks due to the presence of code context, Pearson correlations with other benchmarks reveal different performance trends. This suggests our dataset captures distinct evaluation dimensions, highlighting challenges in environment compatibility and version adaptation that standard benchmarks overlook.
}
Together, these analyses confirm that VersiBCB offers a more realistic and fine-grained assessment of LLM performance in practical software development settings, especially in tasks involving package evolution, compatibility resolution, and cross-version reasoning.

\begin{table}[t]
\centering
\caption{Pearson correlation ($r$) and $p$-value between existing benchmarks and VersiBCB tasks.}
\label{tab:correlation_analysis}
\resizebox{\linewidth}{!}{
\begin{tabular}{@{}lcc@{}}
\toprule
\textbf{Benchmark} & \textbf{VersiBCB-CG} & \textbf{VersiBCB-CM} \\ \midrule
HumanEval+                & $r = 0.925$, $p = 0.700$ & $r = 0.095$, $p = 0.100$ \\
LiveCodeBench (codegen)   & $r = 0.893$, $p = 0.700$ & $r = 0.374$, $p = 0.500$ \\
\bottomrule
\end{tabular}
}
\end{table}

\begin{figure*}[t]
    \centering
    \includegraphics[width=\linewidth]{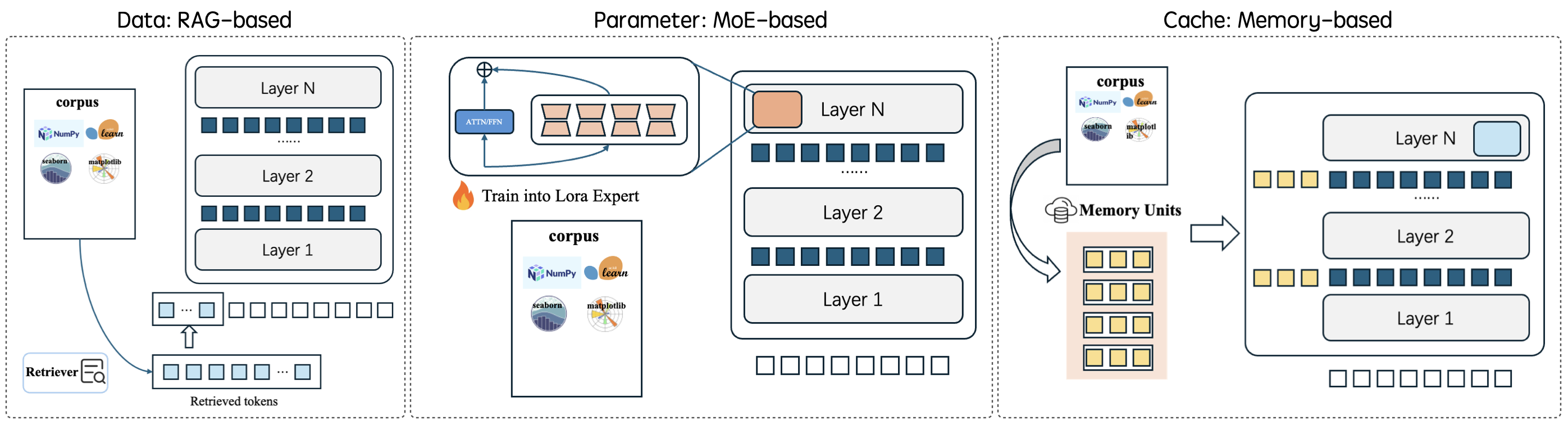}
    \caption{Three axes of environment-aware LLM customization.
    \textbf{Left:} Data-based (RAG) prepends retrieved context.
    \textbf{Middle:} Parameter-based (MoE) activates version-aware experts via gates.
    \textbf{Right:} Cache-based (Memory) injects layerwise key-value pairs learned from prior runs.}
    \Description{Three axes of environment-aware LLM customization.
    \textbf{Left:} Data-based (RAG) prepends retrieved context.
    \textbf{Middle:} Parameter-based (MoE) activates version-aware experts via gates.
    \textbf{Right:} Cache-based (Memory) injects layerwise key-value pairs learned from prior runs.}
    \label{fig:approach}
\end{figure*}

\section{LLM Customization Strategies}
\label{sect:strategy}

Modern software environments rarely share identical dependency stacks, yet large language models are typically deployed as monolithic, one-size-fits-all systems. To address this \emph{environment mismatch}, we explore three orthogonal adaptation axes: \textbf{data}, \textbf{parameters}, and \textbf{state}, where each is formalized in the Transformer view introduced in \S\ref{sect:intro}. Concretely, environment-specific information can be injected at inference time by (i) augmenting the input with retrieved evidence $\mathcal{R}_t$, (ii) dynamically routing hidden states through version-specialized experts using gating weights $g_t^l$, or (iii) pre-loading key–value caches $M_t^l$ that encode environment-conditioned computation patterns. An overview is shown in Figure~\ref{fig:approach}.

\subsection{Data-based Customization: RAG}
\label{subsect:rag}

{\noindent\textbf{Motivation.}
When the context window is available, retrieval augmented generation (RAG) ~\cite{gao2023retrieval} enables lightweight adaptation by injecting fresh, environment-specific artifacts (e.g., API docs, code snippets, deprecation notices). 
This is particularly valuable in EACG scenarios where critical details, such as a missing keyword argument, may only be documented in version-specific changelogs.
Such methods analogous to consulting documentation or examples, injecting environment-relevant context without changing the model.
}


\noindent\textbf{Integration.}
RAG introduces no architectural change; it appends metadata to the input, typically using a lightweight schema such as \texttt{torch==1.9.0}.

\noindent\textbf{Overhead.}
RAG incurs a small token expansion $\mathcal{O}(L_{\text{ret}})$ and external retrieval latency; VRAM usage is negligible.

\noindent\textbf{Strengths and Limitations.}
\emph{Pros:} Immediate adaptation, interpretable evidence, easy deployment.  
\emph{Cons:} Effectiveness degrades with long dependencies; constrained by token window.


\subsection{Parameter-based Customization: MoE}
\label{subsect:moe}

{\noindent\textbf{Motivation.}
Full retraining for every ($L, V$) configuration is infeasible. Instead, we use a \emph{LoRA~\cite{hu2022lora}-based mixture-of-experts} (MoE)~\cite{shazeer2017outrageously} where each expert specializes in a package cluster (e.g., \texttt{torch<1.10}). Gating ensures that only relevant experts are active per request.
Such methods analogous to installing specialized plugins, where model capacity is partitioned into experts aligned with library versions.
}


\noindent\textbf{Integration.}
Each selected expert includes LoRA rank-$r$ adapters; the base model remains frozen.

\noindent\textbf{Overhead.}
Runtime scales by $\sim1.2\times$ for $k=2$; storage cost is $\mathcal{O}(Erd)$ with $d$ as hidden width.

\noindent\textbf{Strengths and Limitations.}
\emph{Pros:} High capacity, task reuse, combinatorial generalization (critical for EACM).  
\emph{Cons:} Training requires supervision; gate collapse risk under data imbalance.

\subsection{Cache-based Customization: Memory}
\label{subsect:memory}

{\noindent\textbf{Motivation.}
Repeated environment patterns (e.g., how \texttt{datasets.\\load\_dataset} is expanded in v1.12) can be pre-cached. Injecting these patterns as prefix-KV enables zero-latency adaptation~\cite{abs-2407-01178}.
Such methods analogous to caching and reusing prior solutions, enabling fast adaptation by re-injecting previously learned environment-specific patterns.
}


\noindent\textbf{Integration.}
Implemented via the standard \texttt{past\_key\_values} argument in decoding APIs.

\noindent\textbf{Overhead.}
VRAM scales with $\mathcal{O}(L_{\texttt{env}} d)$, but no extra FLOPs; highly suitable for latency-sensitive setups.

\noindent\textbf{Strengths and Limitations.}
\emph{Pros:} Instantaneous, stateless, low memory footprint.  
\emph{Cons:} Cannot adapt to unseen configurations (cold-start risk).





\section{Experimental Setup and Results}
\label{sect:experiment}

To evaluate our proposed strategies for environment-aware code generation, we organize our experiments around three core research questions, each targeting a distinct operational challenge in real-world development settings.

\subsection{RQ1: Can the generated or migrated code execute successfully in the specified environment? (Executability)}

\noindent\textbf{Motivation.}
LLMs often produce code that is syntactically correct but fails at runtime due to mismatched APIs or missing dependencies. This RQ evaluates whether environment-aware customization improves executable correctness.

\noindent\textbf{Setup.}
We compare Baseline, RAG, MoE, and Memory variants under held-in configurations using VersiBCB. For a given set of programming tasks, \texttt{pass@1} is the percentage of tasks whose single (top-1) generated solution passes every unit test in its associated suite, offering a strict all-or-nothing measure of correctness. \texttt{wpass@1} relaxes this requirement by first computing, for each task, the proportion of unit tests that the solution passes, and then averaging these proportions across all tasks, thereby providing a continuous score that also credits partial correctness. Pass@1 reflects strict “all-or-nothing” correctness, requiring all unit tests to pass. In contrast, wPass@1 provides a continuous measure that rewards partial correctness, offering a more nuanced view of functional performance.
\begin{align}
\todo{\text{wpass@1} = \frac{1}{\mathcal{D}} \sum_{i=1}^{|\mathcal{D}|} \frac{\text{\# passed tests for } C_i}{\text{total tests for } C_i}}
\end{align}

\begin{table}[t]
\centering
\caption{Pass@1 and wPass@1 performance for EACG/EACM under different customization strategies. Relative gains (\textcolor{gray}{↑}) are also reported.}
\label{tab:exec_result}
\renewcommand{\arraystretch}{1.2}
\resizebox{\linewidth}{!}{
\begin{tabular}{l|c c|c c|c c|c c}
\toprule
\textbf{Method} 
& \multicolumn{4}{c|}{\textbf{EACG (\%)}} 
& \multicolumn{4}{c}{\textbf{EACM (\%)}} \\
 & pass@1 & $\Delta$ & wpass@1 & $\Delta$ & pass@1 & $\Delta$ & wpass@1 & $\Delta$ \\
\midrule
Baseline & 14.63 & -- & 33.16 & -- & 42.51 & -- & 58.94 & -- \\
\midrule
RAG    & 15.22 & \textcolor{gray}{↑0.59} & 34.54 & \textcolor{gray}{↑1.38} & 44.01 & \textcolor{gray}{↑1.50} & 60.81 & \textcolor{gray}{↑1.87} \\
MoE    & 14.93 & \textcolor{gray}{↑0.30} & 39.37 & \textcolor{gray}{↑6.21} & 43.41 & \textcolor{gray}{↑0.90} & 60.48 & \textcolor{gray}{↑1.54} \\
Memory & 15.22 & \textcolor{gray}{↑0.59} & 33.93 & \textcolor{gray}{↑0.77} & 48.20 & \textcolor{gray}{↑5.69} & 63.77 & \textcolor{gray}{↑4.83} \\
\bottomrule
\end{tabular}
}
\end{table}

\noindent\textbf{Results and Findings.} 
We evaluate the executability of code generated under environment-specific constraints using \texttt{pass@1} and \texttt{wpass@1}, comparing three adaptation strategies against the \texttt{LLaMA\\3.1-8B-Instruct} baseline. As shown in Table~\ref{tab:exec_result}, for the code generation task (EACG), the baseline achieves a \texttt{pass@1} of 14.63\% and \texttt{wpass@1} of 33.16\%. RAG and Memory provide marginal gains in \texttt{pass@1} (+0.59\%), while MoE yields the most substantial improvement in \texttt{wpass@1} (+6.21\%), suggesting its ability to capture partial semantic correctness despite limited impact on full executability.
In the code migration task (EACM), where adaptation is grounded in pre-existing context, all methods demonstrate more significant improvements. Memory achieves the highest gains in both \texttt{pass@1} (+5.69\%) and \texttt{wpass@1} (+4.83\%), confirming its effectiveness in leveraging cached environment-conditioned patterns. RAG performs consistently across both tasks, while MoE shows relatively weaker improvements in EACM, potentially due to suboptimal expert activation in unseen migration contexts.

\begin{figure}[t]
    \centering
    \includegraphics[width=0.47\textwidth]{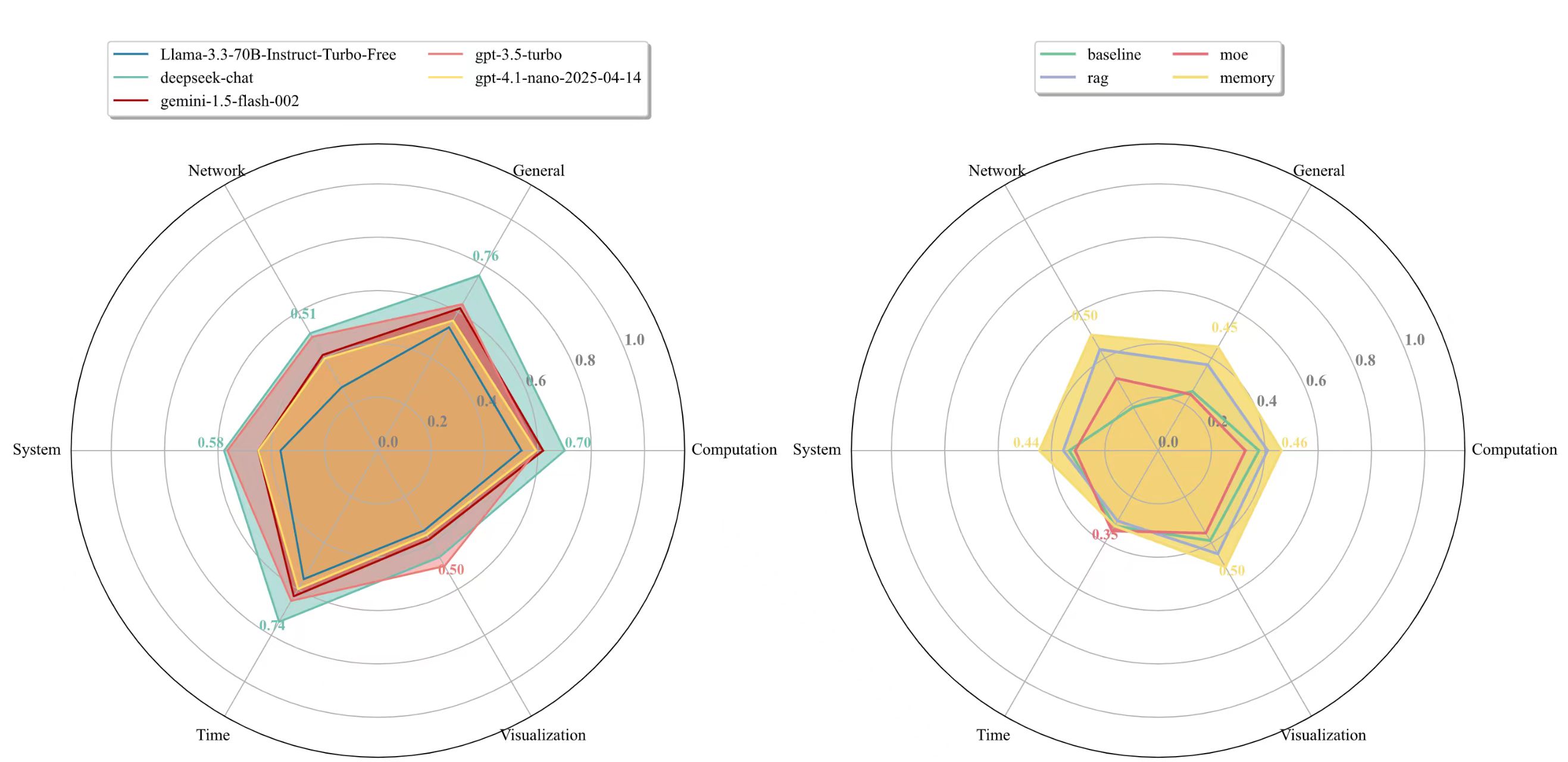}
    \caption{Task-wise Pass@1 across domains (e.g., data processing, computation, ML).}
    \Description{Task-wise Pass@1 across domains (e.g., data processing, computation, ML).}
    \label{fig:exec_domain}
\end{figure}

Figure~\ref{fig:exec_domain} further reveals that performance drops sharply in machine learning scenarios across all methods, highlighting the inherent challenges of adapting to fast-evolving libraries such as \texttt{transformers} and \texttt{torch}, where subtle version changes can severely affect compatibility.

\begin{findingbox}\small
\textbf{Finding 1:} Parameter-based customization (MoE) is most effective at improving partial correctness in code generation, while cache-based memory excels in code migration by reusing environment-specific patterns. However, all strategies exhibit performance degradation in ML-related tasks, indicating a persistent gap in modeling complex, version-sensitive API evolution across domains.
\end{findingbox}



\subsection{RQ2: Are all API calls in the generated code consistent with the specified environment? (Compatibility)}

\noindent\textbf{Motivation.}
Even if code executes, it may use deprecated or incorrect APIs. This RQ examines whether the code conforms to the version-specific library schema.

\noindent\textbf{Setup.}
To rigorously assess the model’s environment awareness in the presence of evolving API landscapes, we curate a controlled subset of the VersiBCB benchmark, consisting of code generation instances annotated with explicit deprecation alerts. Each instance is carefully selected to include both an active API and at least one deprecated but still invocable alternative. This design simulates realistic ambiguity in software maintenance scenarios, where legacy APIs remain callable but are no longer recommended.
To quantify the model’s preference for up-to-date API usage, we report two complementary variants of the pass@1 metric: \texttt{Lenient pass@1}, which treats both active and deprecated APIs as valid, capturing functional correctness regardless of version status. \texttt{Strict pass@1}, which credits only those generations that exclusively invoke active APIs, enforcing alignment with current API schemas and best practices.

\begin{table}[t]
\centering
\caption{Strict and Lenient \texttt{Pass@1} accuracy (\%) on the VersiBCB subset for deprecation-aware code generation (EACG) and code migration (EACM) tasks. \textit{Strict@1} considers only active APIs as correct, while \textit{Lenient@1} also accepts deprecated ones. $\Delta$ denotes the difference between the two, highlighting model sensitivity to deprecation signals.}
\label{tab:deprecation_pass1}
\renewcommand{\arraystretch}{1.2}
\resizebox{.98\linewidth}{!}{
\begin{tabular}{l|ccc|ccc}
\toprule
\multirow{2}{*}{\textbf{Method}} & \multicolumn{3}{c|}{\textbf{EACG}} & \multicolumn{3}{c}{\textbf{EACM}} \\
 & Strict@1 & Lenient@1 & {\textcolor{gray}{$\Delta$}} & Strict@1 & Lenient@1 & {\textcolor{gray}{$\Delta$}} \\
\midrule
Baseline & 21.95 & 30.49 & \textcolor{gray}{↑8.54} & 42.68 & 64.63 & \textcolor{gray}{↑21.95} \\
RAG      & 24.39 & 28.05 & \textcolor{gray}{↑3.66} & 43.90 & 67.07 & \textcolor{gray}{↑23.17} \\
MoE      & 26.83 & 31.71 & \textcolor{gray}{↑4.88} & 50.00 & 80.49 & \textcolor{gray}{↑30.49} \\
Memory   & 23.17 & 32.93 & \textcolor{gray}{↑9.76} & 45.12 & 74.39 & \textcolor{gray}{↑29.27} \\
\bottomrule
\end{tabular}
}
\end{table}

\noindent\textbf{Results and Findings.} 
As shown in Table~\ref{tab:deprecation_pass1}, we report two variants of pass@1: \textit{Strict@1}, which credits only generations using active APIs, and \textit{Lenient@1}, which tolerates both active and deprecated APIs. The gap $\Delta$ between the two reflects a model’s propensity to rely on obsolete interfaces.
For the EACG task, the baseline achieves 21.95\% in Strict@1 and 30.49\% in Lenient@1 ($\Delta$ = +8.54\%), underscoring a substantial reliance on deprecated APIs. Among customized models, the MoE-based variant performs best in Strict@1 (26.83\%) and maintains a modest $\Delta$ (+4.88\%), indicating an improved alignment with version-specific standards. The Memory approach achieves the highest Lenient@1 (32.93\%) but with a wider gap ($\Delta$ = +9.76\%), suggesting its behavior favors executability over strict compliance. RAG offers moderate improvements with the smallest gap (+3.66\%), demonstrating its cautious use of APIs, likely influenced by injected version documentation.
In the EACM setting, all methods exhibit significantly higher scores due to the additional grounding from the input context. MoE again leads in Strict@1 (50.00\%) and achieves a high Lenient@1 (80.49\%), demonstrating strong control over deprecated usage in migration scenarios. Memory exhibits high compatibility (Lenient@1 = 74.39\%) but also the highest reliance on deprecated APIs ($\Delta$ = +29.27\%). RAG offers a more balanced trade-off (Strict@1 = 43.90\%, $\Delta$ = +23.17\%) and outperforms the baseline across metrics.

\begin{findingbox}\small
\textbf{Finding 2:} MoE achieves the highest strict API consistency across tasks, indicating strong alignment with version-specific schemas. Memory improves overall compatibility but tolerates more deprecated usage, while RAG offers the most conservative adaptation. Strict adherence to environmental constraints remains challenging, especially in migration.
\end{findingbox}

\subsection{RQ3: Can the system generalize to unseen library and version combinations? (Composability)}

\noindent\textbf{Motivation.}
A key requirement for scalable deployment is that LLMs generalize to novel configurations not seen during adaptation or fine-tuning.

\noindent\textbf{Setup.}
To rigorously assess the model’s compositional generalization under environment perturbations, we construct a variant of the \texttt{VersiBCB} evaluation suite featuring synthetically perturbed configurations. For each seed instance involving $n$ libraries $\{\ell_1, \ell_2, \dots, \ell_n\}$, we define a \textit{generalization distance} $i \in [1, n]$ as the number of libraries whose versions are randomly altered while keeping the package names fixed. Each perturbed configuration is retained only if both the original reference code and its test suite remain executable under the new environment, ensuring semantic preservation and fair evaluation.
This setup simulates realistic yet unseen environment shifts, where models must maintain functional correctness under partially modified package landscapes. In the baseline setting, version perturbations yield minimal changes to the instruction prompt. In contrast, the customized variants (RAG, MoE, Memory) respond to version shifts by replacing retrieved documentation, rerouting through version-aware experts, or reloading cached key-value states. This explicitly tests the adaptability of each customization mechanism.
We evaluate generalization using two complementary metrics: (i) the average \texttt{Pass@1} over all perturbed instances, and (ii) the Pearson correlation between binary success outcomes and generalization distance. An ideal model should exhibit a weak or non-negative correlation, indicating robustness to environment shifts. A strong negative correlation suggests poor compositional stability and overreliance on memorized patterns.


\noindent\textbf{Results and Findings.}
To evaluate whether environment-aware models can robustly generalize to previously unseen library-version configurations, we synthetically perturb the environments of held-in instances by randomly altering the versions of $i$ libraries ($i \leq 5$) while preserving semantic executability. Table~\ref{tab:composability} reports the average \texttt{pass@1} and the Pearson correlation ($r$) between generation success and perturbation distance.
The {baseline} exhibits limited generalization capacity, with a \texttt{pass@1} of 14.67\% (EACG) and 36.28\% (EACM), accompanied by strong negative correlations ($r = -0.82$ and $r = -0.53$), indicating high sensitivity to version shifts. Among the three customization strategies, both \textbf{RAG} and \textbf{Memory} outperform the baseline in EACG (+1.63\%) but still suffer from severe generalization degradation ($r = -0.84$), suggesting that token-level or cache-level conditioning alone is insufficient for stable adaptation in generation-from-scratch scenarios.

\begin{table}[t]
\centering
\caption{Composability analysis: pass@1 (in \%) and Pearson $r$ under perturbed environments. Values in parentheses show raw counts. $\uparrow$ / $\downarrow$ indicate relative gains/losses compared to baseline.}
\label{tab:composability}
\renewcommand{\arraystretch}{1.2}
\resizebox{\linewidth}{!}{
\begin{tabular}{l|cc|cc}
\toprule
\textbf{Method} & \multicolumn{2}{c|}{\textbf{EACG}} & \multicolumn{2}{c}{\textbf{EACM}} \\
\cline{2-5}
 & pass@1 (\%) & Pearson $r$ & pass@1 (\%) & Pearson $r$ \\
\midrule
Baseline & 14.67 (27/184) & \textcolor{gray}{-0.82} & 36.28 (78/215) & \textcolor{gray}{-0.53} \\
\midrule
RAG      & 16.30 (30/184) $\uparrow$ & \textcolor{gray}{-0.84} & 40.47 (87/215) $\uparrow$ & \textcolor{gray}{1.00} \\
MoE      & 13.59 (25/184) $\downarrow$ & \textcolor{gray}{-0.49} & 33.95 (73/215) $\downarrow$ & \textcolor{gray}{-0.24} \\
Memory   & 16.30 (30/184) $\uparrow$ & \textcolor{gray}{-0.84} & \textbf{45.58} (98/215) $\uparrow$ & \textcolor{gray}{0.85} \\
\bottomrule
\end{tabular}
}
\end{table}


In contrast, in the code migration setting (EACM), \textbf{Memory} yields substantial improvements, with a \texttt{pass@1} of 45.58\% (+9.30\%) and a strong positive correlation ($r = 0.85$), demonstrating that pre-cached environment-specific key-value memories can enhance resilience to compositional environment perturbations. \textbf{MoE}, while architecturally more flexible, shows mixed performance: it underperforms in EACG (-1.08\%), likely due to suboptimal gating under novel conditions, and its compositional robustness under migration is also limited due to gate misrouting. 
Overall, these results highlight the relative advantages of memory-based adaptation in preserving environment-aligned patterns during migration, while exposing persistent challenges in achieving robust compositional generalization in open-ended generation.


\begin{findingbox}\small
\textbf{Finding 3:} Cache-based customization (Memory) yields the strongest compositional robustness in environment-aware code migration, achieving the highest \texttt{pass@1} and stable performance under perturbations. In contrast, MoE is unstable, and RAG lacks resilience. Robust generalization to unseen environment compositions remains an open challenge.
\end{findingbox}




\section{Analysis}

\subsection{Case Studies and Failure Analysis}
\label{sect:case_study}

To complement the quantitative evaluation, we conduct qualitative analyses on representative failure cases to better understand the limitations of environment-aware customization strategies. Our investigation focuses on three orthogonal failure modes, like semantic misalignment, context overfitting, and expert instability, each corresponding to the adaptation mechanism employed.

\subsubsection{RAG Failure: Spurious Context Alignment and Overfitting to Retrieved Evidence}

While Retrieval-Augmented Generation (RAG) provides a lightweight mechanism to inject environment-specific knowledge, we observe systematic failures arising from its excessive reliance on partially relevant context. In one case, although the task clearly specified the need to return both \texttt{skew} and \texttt{kurtosis} statistics, the model erroneously produced two identical invocations of \texttt{scipy.stats.skew}, likely influenced by redundant or ill-structured retrievals that underrepresent the semantic distinction between statistical moments.
Moreover, RAG often fails to respect fine-grained behavioral constraints in downstream requirements. In a visualization task requiring a return type of \texttt{BarContainer}, the model instead generated code that returned an \texttt{Axes} object, conforming to retrieved usage patterns but violating the task specification and test interface. This suggests that, in absence of explicit semantic reconciliation, retrieved snippets can misguide generation even when superficially plausible.

\begin{figure}[t]
\centering
\begin{codebox}\small
\scriptsize
\begin{lstlisting}[style=pycode, language=Python, escapeinside={(*@}{@*)}]
# --- Field ---
import scipy.stats as stats

# --- Correct Code ---
row_max = np.max(matrix, axis=1)
skew = stats.skew(row_max)
kurt = stats.kurtosis(row_max)

# --- RAG-Generated Code ---
row_max = np.max(matrix, axis=1)
skew = stats.skew(row_max)
kurt = stats.skew(row_max)  # (*@\textcolor{red}{ Should be kurtosis}@*)
\end{lstlisting}
\end{codebox}
\caption{Incorrect reuse of \texttt{skew()} generated by RAG. Despite task instruction requiring both \texttt{skew} and \texttt{kurtosis}, RAG overfits retrieved pattern and fails semantic grounding.}
\Description{Incorrect reuse of \texttt{skew()} generated by RAG. Despite task instruction requiring both \texttt{skew} and \texttt{kurtosis}, RAG overfits retrieved pattern and fails semantic grounding.}
\label{fig:kurtosis_rag_fail}
\end{figure}

\begin{findingbox}\small
\textbf{Finding 4.} RAG is susceptible to overfitting retrieved artifacts, resulting in hallucinated or misaligned API usage. The model lacks effective mechanisms to reconcile contextually retrieved evidence with task-specific semantic constraints such as return types and argument structure.
\end{findingbox}

\subsubsection{Memory Failure: Latent Semantics Drift and Cache Interference}

Memory-augmented models aim to improve environment alignment by injecting precomputed key-value pairs encoding API usage patterns. However, such cached information can inadvertently override task-relevant signals when its semantics diverge from current requirements. In one observed failure, the model incorrectly invoked \texttt{rfft} instead of \texttt{fft}, reflecting memorization of patterns suitable for real-valued inputs rather than the complex-spectrum analysis demanded by the task.
Another case involved erroneous parsing of timezone-aware \texttt{datetime} objects: the model misapplied string-based preprocessing logic derived from cached prompts, resulting in type mismatches at runtime. These failures highlight the absence of runtime validation or semantic consistency checks for cache entries, especially detrimental in dynamic or structurally ambiguous domains like date-time processing.

\begin{findingbox}\small
\textbf{Finding 5.} Memory-based customization introduces brittle dependencies on previously cached patterns, which, when semantically misaligned with current task requirements, lead to execution failures. Lack of dynamic cache validation poses significant risks in settings with subtle behavioral constraints.
\end{findingbox}

\subsubsection{MoE Failure: Expert Misrouting and Gate Collapse under Distribution Shift}

The Mixture-of-Experts (MoE) design, grounded in LoRA-based specialization, demonstrates strong potential in enforcing API compatibility. However, its effectiveness depends critically on accurate expert selection. Our case studies reveal that in perturbed environments, particularly those with unseen library-version combinations, the gating mechanism may activate suboptimal experts, resulting in outdated or deprecated API usage.

For example, in a migration scenario involving \texttt{datasets.load\\\_dataset}, the model incorrectly routed input through an expert trained on an older version (1.1.3), producing incompatible arguments for the target environment (1.12.0). This failure mode reflects gate collapse caused by either sparse supervision or non-discriminative routing features, especially in low-frequency or long-tail configurations.

\begin{findingbox}\small
\textbf{Finding 6.} MoE adaptation exhibits instability under environment shifts, where expert gating may activate outdated or irrelevant modules. This reveals a core limitation in scaling fine-grained specialization without robust routing mechanisms or broader expert coverage.
\end{findingbox}

\subsection{Scaling Analysis for Parameterized Training and Inference}

While parameter-based customization (e.g., LoRA-based Mixture-of-Experts) demonstrates notable improvements in environment-aware generation tasks, its deployment in real-world scenarios requires a thorough assessment of scalability and transferability. Unlike stateless strategies such as retrieval or memory caching, MoE introduces additional training and storage overheads due to the need for specialized expert modules tied to environment configurations. This section aims to quantify these trade-offs by addressing two fundamental questions: \textbf{(Q1)} To what extent can LoRA experts trained on a base model (e.g., \texttt{LLaMA3.1-8B}) be reused on its instruction-tuned variant (\texttt{LLaMA3.1-8B-Instruct}) without degradation? \textbf{(Q2)} How does MoE performance scale with the capacity of the underlying base model (e.g., 7B, 13B, 34B), assuming consistent architecture?

\subsubsection{Cross-Variant Transferability of LoRA Experts.}
The cost of training and maintaining environment-specific experts becomes prohibitive if each base model variant (e.g., instruction-tuned) requires isolated fine-tuning. Assessing whether LoRA modules are reusable across model variants with shared architecture but divergent training objectives is critical for practical scalability.

\noindent\textbf{Experimental Setup.} We evaluate expert transferability by training LoRA experts under \texttt{LLaMA3.1-8B}, and applying them, without further tuning, to its instruction-tuned variant, \texttt{LLaMA3.1-8B-\\Instruct}. Results are benchmarked against experts natively trained on the target model.

\noindent\textbf{Results.} Table~\ref{tab:adapter-transfer} shows that transferred experts maintain comparable performance on both EACG and EACM tasks, achieving similar pass@1 and wpass@1 scores. Notably, while the instruct-trained expert slightly outperforms in EACG, the transferred expert yields higher EACM success, suggesting resilience across related variants.

\begin{table}[t]
\centering
\caption{\todo{
Performance of cross-variant adaptor reuse and in-distribution tuning on VersiBCB. 
InstAdaptor refers to experts trained on the instruction-tuned model (\texttt{LLaMA-3.1-8B-Instruct}), whereas BaseAdaptor denotes experts trained on the base model (\texttt{LLaMA-3.1-8B}) and reused on the instruction-tuned variant without further tuning.
}}
\label{tab:adapter-transfer}
\renewcommand{\arraystretch}{1.2}
\resizebox{\linewidth}{!}{
\begin{tabular}{l|l|cc|cc}
\toprule
\textbf{Model} & \textbf{Approach} & \multicolumn{2}{c|}{\textbf{EACG}} & \multicolumn{2}{c}{\textbf{EACM}} \\
& & pass@1 & wpass@1 & pass@1 & wpass@1 \\
\midrule
\multirow{3}{*}{LLaMA-3.1-8B-Instruct} 
& Baseline (w/o Adaptor) & 14.63 & 33.16 & 42.51 & 58.93 \\
& MoE (w. InstAdaptor)   & \textbf{15.22} & 39.10 & 40.42 & 57.55 \\
& MoE (w. BaseAdaptor)   & 14.93 & \textbf{39.40} & \textbf{43.41} & \textbf{60.48} \\
\bottomrule
\end{tabular}}
\end{table}

\begin{findingbox}\small
\textbf{Finding 7.} LoRA experts trained on base models exhibit strong transferability to structurally aligned instruction-tuned variants, mitigating the need for redundant fine-tuning across closely related checkpoints.
\end{findingbox}

\subsubsection{Scalability Across Model Sizes.}

In high-stakes software environments, adopting larger language models is often desirable for improved reasoning capacity and robustness. We investigate whether parameter-based environment adaptation remains effective as model capacity scales up, a necessary condition for broader deployment in diverse development stacks.

\noindent\textbf{Experimental Setup.} We apply MoE to \texttt{CodeLLaMA} models of increasing scale (7B, 13B, 34B), training separate LoRA experts per configuration under equivalent adaptation procedures. Performance on EACG and EACM is measured relative to the 7B baseline.

\begin{figure}[t]
    \centering
    \includegraphics[width=\linewidth]{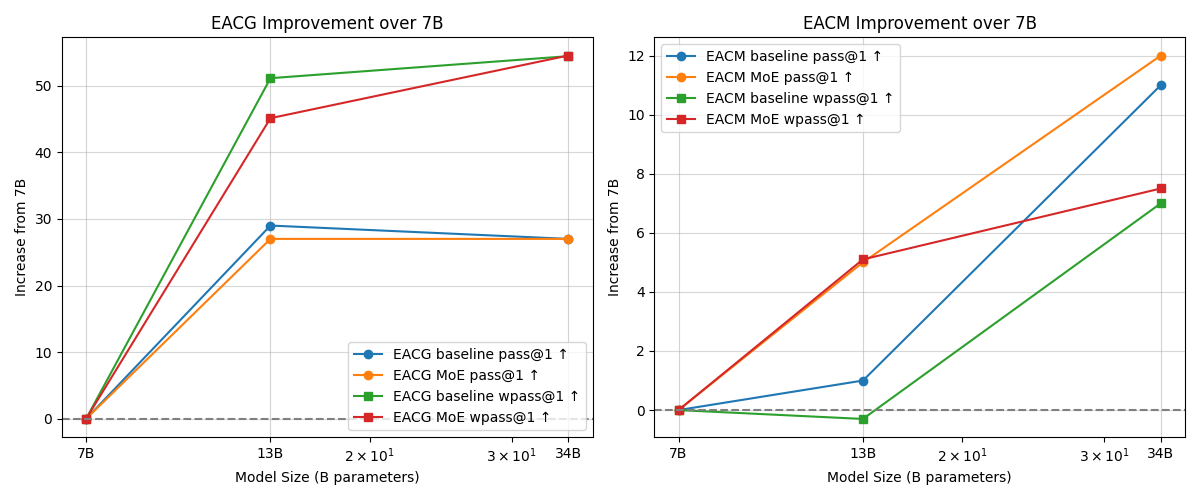}
    \caption{Scaling behavior of MoE-based environment customization across CodeLLaMA-7B/13B/34B.}
    \Description{Scaling behavior of MoE-based environment customization across CodeLLaMA-7B/13B/34B.}
    \label{fig:scaling}
\end{figure}

\noindent\textbf{Results.} As illustrated in Figure~\ref{fig:scaling}, we observe a monotonic improvement in both code generation and migration quality with larger model sizes. The largest model (34B) yields substantial gains, indicating that MoE benefits scale effectively with capacity under the same architectural regime.

\begin{findingbox}\small
\textbf{Finding 8.} Parameter-based customization via MoE exhibits favorable scaling behavior: as model size increases, environment-aligned experts yield consistent and amplified gains in both EACG and EACM performance.
\end{findingbox}


\section{Discussion}
\label{sect:deployment}

While our customization strategies (RAG, MoE, and Memory) have shown tangible improvements in aligning LLMs with environment-specific requirements, several critical challenges remain before such models can be seamlessly integrated into production workflows.

\noindent\textbf{Environmental Robustness Remains the Bottleneck.}\todo{
Even the strongest proprietary model we evaluated (GPT 4.1) reaches only 27.76 Pass@1 on our benchmark (Table~\ref{tab:model_performance}), well below practical deployment thresholds. A preliminary error analysis shows that its failure rate increases by roughly 40\% on unknown versions released after its pretraining window, revealing substantial temporal brittleness. This temporal brittleness highlights the fragility of today’s models under realistic dependency drift and underscores the need for targeted adaptation rather than static improvements on fixed datasets. Our focus, therefore, is not on surpassing proprietary models, but on understanding where they fail and identifying adaptation levers capable of closing the reliability gap.}


\noindent\textbf{Deployment Requires Rethinking Integration and Workflow Design.}
Practical adoption depends on how environment aware capabilities integrate into development workflows. Plugin style strategies such as RAG and memory offer low latency, IDE level support but are constrained by context length and local compute. Service oriented approaches such as MoE scale expertise centrally but introduce challenges related to orchestration, privacy, and latency. Real integration further requires standardized interfaces for environment introspection, reliable context injection, and fallback mechanisms when dependency metadata is incomplete. 

\noindent\textbf{Toward Continual, Software Aligned LLMs.}
Environment aware code generation is inherently dynamic: package versions evolve, APIs deprecate, and configurations compose in combinatorial ways. Robust generalization under such change likely requires models that are continually updated rather than statically adapted. Promising directions include changelog aware fine tuning, dynamic routing guided by online performance feedback, and extending environment awareness beyond Python packages to container specifications, hardware constraints, and operating system dependencies. Advancing these directions will strengthen the reliability of software aligned reasoning.

\section{Related Work}
\label{sec:related}

\noindent\textbf{Environment-Aware Code Generation.}
Recent advancements in LLM-based code generation~\cite{chen2021evaluating, lai2023ds, gpt4o,oneeval} have demonstrated strong performance on standard benchmarks, yet these models are typically evaluated under default or implicit environments, ignoring the practical challenges posed by version-specific dependencies~\cite{wu2024versicode,xinp_evo}. LibEvolutionEval~\cite{libevolutioneval2025} reveals performance degradation across library versions but lacks execution-grounded assessment or environment conditioning. Versicode~\cite{wu2024versicode} introduces symbolic version control for API calls, but it does not ensure executability nor consider cross-library interactions. Migration-oriented datasets such as GitChameleon~\cite{abs-2411-05830} and CodeUpdateArena~\cite{chen2024apigen} focus on single-library updates and synthetic settings, offering limited support for realistic, multi-library version transitions. In contrast, our work introduces VersiBCB, a large-scale benchmark that systematically captures real-world Python environments, supports both generation and migration tasks, and incorporates execution-based validation. This enables fine-grained evaluation of LLM behavior under compositional version constraints with API deprecation and cross-version compatibility.

\noindent\textbf{LLM Customization for Software Contexts.}
To adapt general-purpose LLMs to specific downstream contexts, prior work~\cite{lifespan0170HWHZSCHWH0M25} explores strategies across data, parameter, and memory axes. 
Retrieval-Augmented Generation (RAG)~\cite{abs-2405-02355,abs-2312-10997,min_RAG} enhances input prompts with external evidence but is sensitive to retrieval noise and context window limitations. Parameter-based approaches~\cite{HanGL0Z24}, including LoRA~\cite{HuSWALWWC22} and sparse MoE~\cite{MOECaiJWTKH25}, enable scalable specialization but face challenges in expert routing under unseen distributions. Cache-based mechanisms~\cite{abs-2407-01178,cacheabs-2503-10714}, while widely adopted for efficiency, have not been systematically leveraged for environment-aware reasoning. Existing evaluations of these strategies are typically conducted in static, domain-specific tasks and do not account for the combinatorial variability of real-world software environments~\cite{oneeval,recode}. Our work is the first to comparatively assess these adaptation strategies in the context of environment-aware code generation and migration, offering a unified framework to measure their impact on executability, compatibility, and compositional generalization.

\section{Threats to Validity}

\textbf{Threats in benchmark construction.} 
A primary threat in constructing VersiBCB is potential overlap between benchmark tasks and the training data of large language models, which could lead to overestimated performance. 
To mitigate this, VersiBCB is built from expert-curated, execution-verified migration tasks that are grounded in real-world projects but are carefully checked to avoid direct duplicates with common pretraining datasets.
Another threat concerns the subjectiveness in defining environment transitions and labeling code migrations. 
We address this by relying on agent-assisted repair, multi-source documentation, and consistent validation through execution-based checks.
\todo{Note that, all final collected samples have been manually validated to ensure correctness and reliability.}
Moreover, the current focus on Python projects and selected libraries may limit the generalizability of our findings to other languages or package ecosystems, though VersiBCB spans diverse domains and complex multi-library scenarios. 

\noindent\textbf{Threats in empirical evaluation.} A primary threat in our empirical evaluation is that model performance may be affected by implementation bugs, dependency versions, or prompt selection. To mitigate these risks, we adopt public model releases following official guidelines, carefully curate environment specifications, and conduct pilot studies to select effective prompt templates.
Another threat is the potential for interference or inconsistency arising from environmental factors during evaluation. To address this, all experiments are run in isolated environments that strictly match the specified dependencies, ensuring a fair comparison. 
Our use of synthetic environment perturbations and agent-assisted repair presents an additional threat, as these approaches may not fully capture the complexity of real-world software systems.
While our analyses provide insight into compositional generalization, further work is needed to evaluate models on adversarial or previously unseen environments.





\section{Conclusion}
\label{sect:conclusion}

This work makes the first attempt to systematically evaluate LLMs on environment-aware code generation. We introduce VersiBCB, a large-scale benchmark featuring expert-curated, execution-verified migration and generation tasks from real Python projects. Our study of three representative LLM adaptation strategies: retrieval-augmented generation, LoRA-based mixture-of-experts, and memory-augmented generation, reveals that while these methods can improve executability and compatibility under explicit environment constraints, all models still face significant challenges in complex domains and when generalizing to unseen package and version combinations. These findings highlight a persistent gap between current LLM capabilities and the practical requirements of real-world software development, and call for further research on more robust, adaptive environment-aware code generation.
Future efforts may explore integrating runtime environment introspection and container-level profiling into LLM inference pipelines, enabling on-the-fly adaptation to system-level constraints beyond package versions.
Additionally, extending environment-aware code generation to other ecosystems such as Java or Dockerized workflows would further promote broader adoption across enterprise-grade development stacks.


\begin{acks}
    This research was supported by eBay Inc.
\end{acks}

\bibliographystyle{ACM-Reference-Format}
\bibliography{bibfile}

\end{document}